\documentclass[12pt]{iopart}

 \usepackage{iopams} 
 \usepackage{amssymb}
\usepackage{graphicx}
\usepackage{epstopdf}

\begin{document}

\title[]{The stability and energy exchange mechanism of divergent states with real energy}

\author{Hao Jiang, Xiang-Jun Kong and Hui-Ping Huang}
\address{School of Physics and Optoelectronics, Xiangtan University, Xiangtan 411105, China}
\ead{hjiang@xtu.edu.cn}



\vspace{10pt}
\begin{indented}
\item[Keywords: ] non-hermitian, open boundary condition, divergent state, energy density.
\end{indented}

\begin{abstract}
The eigenvalue of the hermitic Hamiltonian is real undoubtedly. Actually, The reality can also be guaranteed by the $PT$-symmetry. The hermiticity and the $PT$-symmetric quantum theory both have requirements regarding the boundary condition. There exists a reverse strategy to investigate the quantum problem. Namely, define the eigenvalue as real first, and, meanwhile, open the boundary condition. Then the behaviors of the wave function at the boundary become rich in meaning. This eigenfunction is generally divergent, and the extent and direction of divergence are closely linked to the energy. It was noted that these divergent behaviors can be well described by their energy-space uncertainty relation which is not trivial anymore. The divergent state is unstable and will certainly exchange energy with the outside. The mechanism of energy exchange is just in the energy-space uncertainty relation, which will benefit dynamic simulation, the many-body problem, and so on. There is no distinct dividing line between this kind of divergent unstable state and the convergent stable state. Their relationship is like that of the rational and irrational numbers. In practice, there are distinct advantages of speed and accuracy for the methods based on the laws of divergence.
\end{abstract}

%
%
%
%
%

\section{Introduction}
The physical observable should be real. In conventional quantum mechanics, the reality of the eigenvalue of a Hamiltonian is guaranteed by its hermitian that comprises the transposition and complex conjugate. If the Hamiltonian contains a differential term, its hermiticity depends on the boundary condition\cite{Maya-MendietaOliveros-Oliveros-201}. In $PT$-symmetric quantum theory which is viewed as a complex generalization of conventional quantum mechanics, the transposition and complex conjugate are replaced by the parity $P$(space reflection) and time reversal $T$\cite{Bender-199}. As long as the Hamiltonian is $PT$-symmetric, its energy levels are in the real domain\cite{BenderBrody-182,BenderMeisinger-212}. $PT$-symmetric, which is not associated with the boundary condition directly, is a weaker restriction than the hermitian. Yet, the reality of the energy is also closely related to the boundary condition\cite{BenderM-232,Bender-185,Mostafazadeh-228}. As noted by Bender, Dirac Hermiticity is too restrictive\cite{Bender-199}. Non-Hermitian quantum theory have a host of advantages\cite{Eec-179,BenderBrody-181}. It has been applied to various fields, such as resonance states, waveguides, dynamics and so on\cite{El-GanainyMakris-229,Eec-179,Nesterov-197,GraefeKorsch-222,IbanezNezgaraot-195,Olendski-146}. The boundary condition is vital to the reality of the Hamiltonian both in conventional quantum mechanics and Non-Hermitian quantum mechanics. Morsy and Ata developed a method to hide the boundary condition in a modified Hamiltonian, which is called the intrinsically Hermitic operators\cite{MorsyAta-214,Maya-MendietaOliveros-Oliveros-201}. As noted above, the customary method is to control the boundary condition to obtain a real number of expectation. If the eigenvalue is defined as a real number at first, and release the boundary condition, then what does the wave function at the boundary look like?  

In this paper, we will do the opposite. Consider a particle in a 1D arbitrary potential well, and the eigenvalue of the Hamiltonian is defined as a real value first, meanwhile open boundary condition is adopted, then the behaviors of the wave function under an open or semi-open boundary condition (semi-open means one side is open, the other side is not) is studied. Generally, this kind of eigenstate is divergent. The divergence shows obvious regularities. The relation between the divergence and the eigenenergy was studied. The uncertainty of the energy is relevant to the stability of the divergent state, while the stability is relevant to the extent of divergence. Therefore, the three of them are interrelated. We found that the divergence that occurs here can be well interpreted via the energy-space uncertainty relation which becomes non-trivial. The divergent state is unstable, and will exchange energy with the outside. The mechanism of energy exchange is investigated through the energy-space uncertainty relation. A strategy is presented to normalize the divergent state without complex scaling\cite{Reinhardt-219,Moiseyev-221}. Divergent phenomena are not unusual in quantum theory, however, their interpretation is rare\cite{HatanoKawamoto-69,Madrid-107}. We will explore the essence of the divergence occurred in the quantum state with real energy. 

\section{Method}
To study the special wave function that is unconstrained or partially constrained by the boundary condition, we combined the transfer matrix method\cite{Johnson-91,VigneronLambin-92,GhatakThyagarajan-94} and the shooting method\cite{GiordanoNakanishi-32,PressTeukolsky-33,Harrison-42}. Considering a segmented 1D potential, the solution for the time-independent Schr\"odinger equation on a certain segment is the superposition of a plane wave. That is, $\Psi_i=A_ie^{-\mathrm{i}k_ix}+B_ie^{\mathrm{i}k_ix}$, where $k_i=\sqrt {2m(E-V_i)}$. A natural unit is adopted in this work, namely, $\hbar=1$, mass $m=1$ and so on. According to the continuity condition, there are two equations at border $x_i$, namely $\Psi_i(x_i)=\Psi_{i+1}(x_i)$ and $\Psi_i'(x_i)=\Psi_{i+1}'(x_i)$. The transfer relations can be obtained by solving the set of equations\cite{Jirauschek-2,JonssonEng-7}. 

At each border, there are two independent equations, among which there are five unknown quantities $A_i$, $B_i$, $A_{i+1}$, $B_{i+1}$ and $E$. Under normal circumstances, $A_i$ and $B_i$ in expression $\Psi_i$ are unknown. Since the wave function multiplied by a nonzero number does not change its eigenvalue, when choosing $x_0$ as the transfer starting point, it is safe to set $A_0=1$ (except for the special situation where $x_0=+\infty$ when a semi-open boundary condition is employed). To obtain the eigenstate at specific energy $E$: 1) regarding the open boundary condition, the transfer starting point is totally free. $B_0$ could be assigned to an arbitrary value, which could be adjusted to an appropriate value by tuning the divergent extent of the wave function. 2) Regarding the semi-open boundary condition, the transfer starts at the boundary that is not open, and then $B_0$ is knowable through the semi-open boundary condition, such as in a bounded situation and $x_0=-\infty$, so we have $B_0=0$. For convenience, semi-open boundary conditions are mainly employed in this work. Then there are only two variates $A_{1}$ and $B_{1}$ at the border of $x_0$ and $x_1$. Continuing on, we can directly obtain the wave function $\Psi_E(A_i, B_i)$ from the transfer relations. There is no doubt that $\Psi_E$ satisfies $\hat{H}\Psi_E=E\Psi_E$. Namely, these solutions are energy eigen-states without exception. Note that the number of eigen-states of a certain energy is countless under an open boundary condition, but for a semi-open boundary condition the corresponding eigenstate is only one except for degeneration which will not appear in 1D. The conclusions of this work are the same as for the open boundary condition if the most stable state is chosen. For convenience and conciseness, the semi-open boundary condition has been adopted.

\section{Physical Laws in a Divergent State}
\begin{figure}
\begin{center}
\includegraphics[width=4.5in]{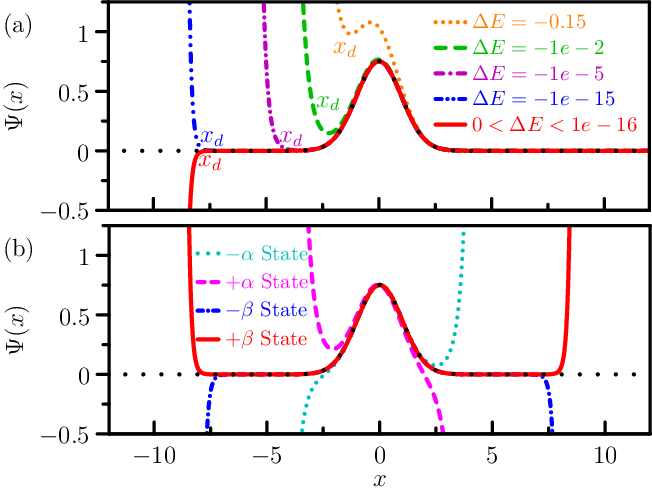}
\caption{\textbf{Wave functions with arbitrary energy.} The eigen-states of a 1D equably segmented harmonic oscillator ($V=\frac 12 x^2$) under open and semi-open boundary conditions are shown. The black dotted line is the analytical solution of the ground state. $\Delta{E}=E-E_0$. (a) Transfer starts at $x=+\infty$ with a semi-open boundary condition. Here, the wave functions are normalized. The positions labeled by $x_d$ are related to the area where the wave function diverges fast. (b) Transfer start at $x=0$ with an open boundary condition.}\label{f1}
\end{center}
\end{figure}

Figure \ref{f1} shows the wave functions $\Psi(x)$ obtained by the above method. They can be classified into two groups. $\alpha$ states: when $\Psi({-\infty})=+\infty$, and $\Psi({\infty})=-\infty$, or when $\Psi({-\infty})=-\infty$, and $\Psi({\infty})=+\infty$, such as the cyan and magenta line in figure \ref{f1} (b). $\beta$ states: when $\Psi({-\infty})=+\infty$, and $\Psi({\infty})=+\infty$, or when $\Psi({-\infty})=-\infty$, and $\Psi({\infty})=-\infty$ ,such as the blue and red line in figure \ref{f1} (b). $\pm\beta$ could be grouped in detail according to the divergence direction. When the transfer starts at $x=\pm\infty$, all states obtained belong to the $\beta$ states, such as the states in figure \ref{f1} (a). It is noticeable that the diverging direction and extent are related to the sign of $\Delta{E}$ and the absolute value of $\Delta{E}$, respectively. For the above regularities of divergence, we surmise that this kind of divergent state should have its own meaning. Actually, divergent phenomena also emerge in the resonant state\cite{Simon-227,Hatano-73,Peierls-89,Gamow-37,TaylorNazaroff-47,Feshbach-48}, but such regularities does not appear in the complex energy state or are deeply hidden. To clarify the regularities of divergence that happen in a semi-open boundary condition, there are at least three questions that should be carefully considered. (\romannumeral1) what does the position $x_d$ denote? (\romannumeral2) how do we interpret the divergence in the wave function? (\romannumeral3) what is the difference between two different diverging directions?

\subsection{What does the position $x_d$ denote?}
In figure \ref{f1} (a), the closer to ground state the energy $E$ is, the further to the left $x_d$ moves. Similar results also exist around excited states. As for an arbitrary energy, the corresponding eigenstate is not always stable. It implies that the energy here also shows uncertainty. Thus, there should be some kind of uncertainty relation between energy $E$ and $x_d$ because the Hamiltonian is no longer hermitic, and the wave function is divergent. The commutation relation cannot be calculated normally. Fortunately, the non-hermitian commutation relation of energy and space is the same as the normal commutation relation\cite{Maya-MendietaOliveros-Oliveros-201}. The definition of uncertainty of a divergent state no longer agrees in the usual sense\cite{Zeldovic-70}. The expression of the uncertainty relation between $\Delta{E}$ and $\Delta{x}$ is $\Delta{E}\Delta{x}\ge {\frac {\hbar}{2m}} |\overline{p}|$. There is no problem in interpreting it with probability at a convergent state. Yet, in a divergent state, $\Delta{E}\neq0$, and generally, $\overline{p}\neq0$. Through comparative analysis, it is easy to determine that the divergent state is unstable. The uncertainty of an unstable state does not agree with the conventional meaning. Since a divergent state is unstable, the most reasonable expression of $\Delta{E}$ is $\Delta{E}\approx|E-E_n|$, where $E$ denotes arbitrary energy, and $E_n$ denotes the energy of the nearest convergent stable state. Through the uncertainty relation, we know that $\Delta{x}$ is inversely proportional to $\Delta{E}$. In figure \ref{f1} (a) the left side wave functions of $x_d$ are the divergent part(DP), and the right side wave functions of $x_d$ coincide with the conventional state, namely, the normal part(NP). Note that the larger $\Delta{E}$ is, the smaller $|x_d|$ is. This leads us to consider that $\Delta{x}$ is potentially related to $|x_d|$. Actually, the exact position of $x_d$ is unclear, which might be a manifestation of the uncertainty principle. We choose the position where the derivative of the wave function is big enough to be $|x_d|$.

In order to confirm the definition of the members in the uncertainty relation: $|\Delta{E}\Delta{x}|\ge {\frac {\hbar}{2m}} |\overline{p}|$, we collected the values of the corresponding $x_d$ of different $\Delta{E}$. We will fit these data in the following.

The two biggest difficulties in fitting are that the mean momentum $\overline{p}$ changes with energy $E$ and the normalization of a divergent wave function. $\langle \Psi|\hat{p}|\Psi \rangle$ cannot get the result directly, since $|\Psi \rangle$ is divergent and unstable. In addition, the divergent part of the wave function cannot be interpreted by probability directly. We will start by promoting an existing concept that is well-known in quantum optics\cite{GerryKnight-267,BarnettRadmore-268,Cohen-TannoudjiDupont-Roc-269,ScullyZubairy-270}. The operator of the total energy of electromagnetic field is
\begin{equation}
\hat{H}=\frac 1 {8\pi}\int ({\bf{}} \hat{\bf{E}}^{2} + \hat{\bf{B}}^2) \rm d\upsilon \qquad(CGS). \label{e2}
\end{equation}
After integration, $\hat{H}$ can be expressed by particle number representation, namely
\begin{equation}
\hat{H}=\sum_{j}\hbar\omega_j(a_j^{+}a_j+\frac 12).
\end{equation}
Consider a single-model field, $\hat{H}=\hbar\omega(a^{+}a+\frac 12)$. The corresponding eigenfunction is $\Psi(q)$, where $q=C(a+a^{+})$. Now, we will find the exact meaning of $\Psi^*(q)\hat{H}\Psi(q)$ which could be interpreted as the energy density at first sight. That is,
\begin{equation}
\Psi^*(q)\hat{H}\Psi(q)=\frac 1 {8\pi}\Psi^*\int ({\bf{}} \hat{\bf{E}}^{2} + \hat{\bf{B}}^2) \rm d\upsilon \Psi.\label{e4}
\end{equation}
Through Fourier transform, we have $\Psi(q)=\int \psi(\textbf{r}) e^{-i q r} \rm d\upsilon$, so equation (\ref{e4}) can be written as

\begin{equation}
\eqalign{\Psi^*(q)\hat{H}\Psi(q)&=\frac 1 {8\pi}\int {\bf{}}\psi^*(\textbf{r})( \hat{\bf{E}}^{2} + \hat{\bf{B}}^2) \psi(\textbf{r})\rm d\upsilon \cr
&= \langle{\bf{}}\psi(\textbf{r})| \frac 1 {8\pi}(\hat{\bf{E}}^{2} + \hat{\bf{B}}^2) |\psi(\textbf{r})\rangle \cr
&=\overline{\frac 1 {8\pi}(\bf{E}^{2} + \bf{B}^2)} \ . \label{e5} }
\end{equation}

From equation (\ref{e5}), we know that $\Psi^*(q)\hat{H}\Psi(q)$ is definitely the energy density but averaged in real space. $\Psi^*(q)\hat{H}\Psi(q)$ is a function of $q$. Replace the representation so that $\psi^*(\textbf{r})\hat{H}\psi(\textbf{r})$ denotes the density of energy distribution averaged in $q$, which is a function of $\textbf{r}$. The above derivation process is based on the electromagnetic field, whose energy is distributed in space, so the concept of energy density is naturally without doubt. Although more direct evidence is needed to generalize it to a real particle, we assume that the concept of density of energy distribution is also applicable to a real particle. Its feasibility will be tested in the following. 

If an eigenstate has been normalized by
\begin{equation}
\eqalign{\langle \Psi(\textbf{r})|\Psi(\textbf{r}) \rangle&=\langle \phi(\textbf{r})|\hat{H}|\phi(\textbf{r})\rangle \cr
&=\langle \phi(\textbf{r})|\phi(\textbf{r})\rangle E \cr
&=E \;,}\label{e6}
\end{equation}
then $\Psi^*(\textbf{r})\Psi(\textbf{r})$ is the density of energy distribution in real space, which is averaged in another vector space. In equation (\ref{e6}), $\langle \phi(\textbf{r})|\phi(\textbf{r})\rangle=1$ is the expression of probability normalization, and $\langle \Psi(\textbf{r})|\Psi(\textbf{r}) \rangle=E$ could be unscrambled by the energy conservation principle. Equations (\ref{e5}) and (\ref{e6}) indicate that probability is possibly related to energy distribution. Therefore, $\Psi$ can be regarded as the superposition of the eigenstate of energy density which differs from the energy eigenstate.

Taken together, the normalization principle seems like another form of the energy conservation principle, and the probability distribution is proportional to the energy distribution. If we know the energy distribution, we, in turn, know the probability distribution. The divergent part denotes instability, so under the concept of energy distribution, the energy distributed at the divergent part should be $\Delta{E}$. (\romannumeral1) Regarding $E_n<E<E_m$, where $E_m\approx \frac{1}{2}(E_n+E_{n+1})$, the total energy distribution is $E$. $E=E_n+\Delta{E}$, so the energy distribution in the normal part is $E_n$, and the corresponding probability is $\frac {E_n}{E}$. Then, the total probability of the divergent part is $\frac {\Delta{E}}{E}$. (\romannumeral2) Regarding $E_m<E<E_n$, where $E_m\approx \frac{1}{2}(E_{n-1}+E_n)$, $\Delta{E}<0$, but we know that the real existing energy should be greater than zero in this system. Thus, the divergent part does not distribute energy, but it denotes the value of the energy it is going to absorb. Note that the state is unstable. The total energy $E$ is all distributed to the normal part. Actually, the probability must be a positive value. Before normalization, the total probability was the sum of the normal part and the divergent part, that is, $|E-E_n|+E=E_n$. Thus, the total probability of the divergent part is $\frac {|\Delta{E}|}{E_n}$.

The local wave functions at a narrow enough interval of the divergent part are approximate momentum eigen-states, because one of the coefficients $A_i$, $B_i$ of this part is infinitely small. Finally, the mean momentum of the divergent part can be expressed by $\overline{p}_d=a\frac {\Delta{E}}{E}|k_i|$ or $\overline{p}_d=a\frac {\Delta{E}}{E_n}|k_i|$, where $a$ is a coefficient imported by the choosing of $|k|$ and is kept unchanged for the different energy $E$. The mean momentum of the normal part should be calculated by $\langle \Phi_{NP}|\hat{p}|\Phi_{NP} \rangle$, and $\Phi$ should be normalized first. Again, with the help of the energy distribution, the normalization process can be achieved easily by $\langle \Psi_{NP}|\Psi_{NP} \rangle E=E_n$($E_n<E<E_m$) or $\langle \Psi_{NP}|\Psi_{NP} \rangle E_n=E$($E_m<E<E_n$). Finally, we have $\overline{p}=\overline{p}_d+\overline{p}_c$.
\begin{figure}
\begin{center}
\includegraphics [width=4.5in]{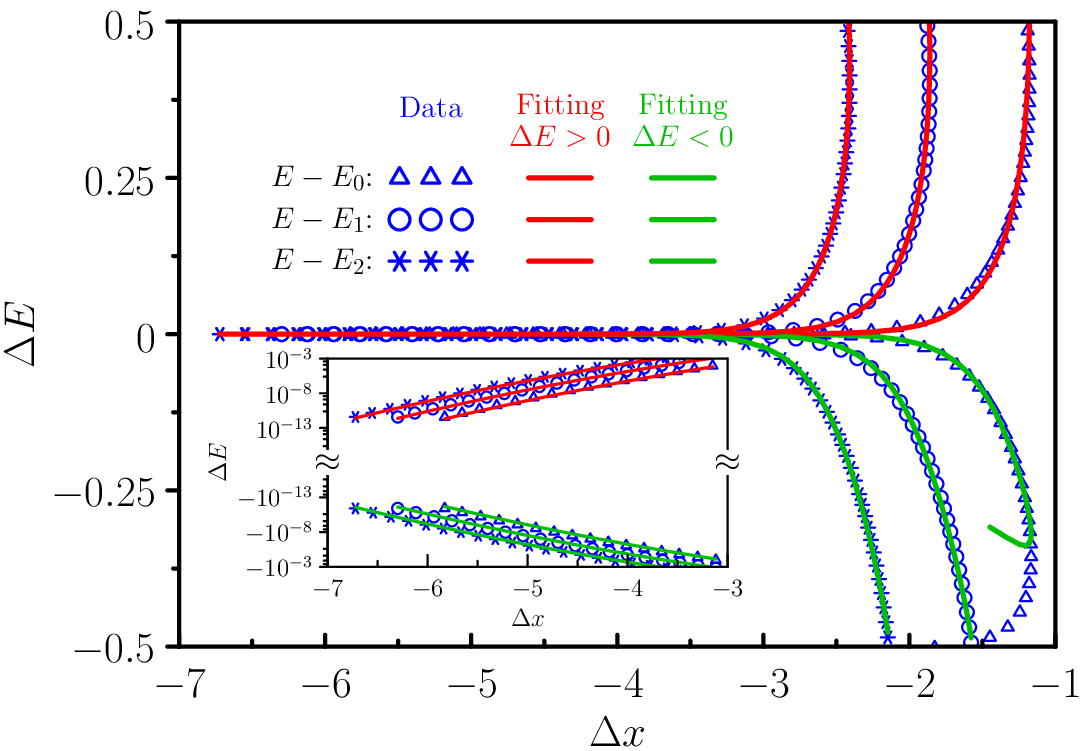}
\caption{\textbf{Fitting results.} The fitting data are collected from the normalized divergent states of the segmented harmonic oscillator. The fitting formula is $\Delta{E}= \frac {\overline{p}}{2 |\Delta{x}|}$. The small inner figure shows the fitting results in a logarithmic pattern.}\label{figdEdx}
\end{center}
\end{figure}

Up to now, the expression of the uncertainty relation could be written as $|\Delta{E}|\ge \frac {|\overline{p}_d+\overline{p}_c|}{2|\Delta{x}|}$. The results of fitting are plotted in figure \ref{figdEdx}. First, the fitting curves match the simulation data well even when $|\Delta{E}|$ is extremely small, such as in the logarithmic plot in figure \ref{figdEdx}. Second, for a certain value of $\Delta{x}$, $|\Delta{E}|$ of the lower energy level $E_n$ is smaller, which implies that the lower energy level is more stable. Third, the position of $x_d$ has a remarkable effect on the fitting beause $\overline{p}_c$ depends on $x_d$. Through simple mathematical derivation, one can get $|\overline{p}_c|=\frac1 2 \phi^2(x_d)$. Generally speaking, similar to figure \ref{f1} (a), the smaller $|x_d|$ is, the bigger the $\phi(x_d)$ is. Therefore, when $|x_d|$ is small, the error in $x_d$ has a remarkable effect on $\frac1 2 \phi^2(x_d)$. This situation is prominent in $E<E_0$ which visibly emerged in the bottom right corner of figure \ref{figdEdx}. The above result is not case-by-case, as the 1D hydrogen atom, limited depth potential well, and arbitrary potential well have similar results. The regularities in divergence have been well described by the energy-space uncertainty relation, so the interpretation of the energy distribution of a particle has been proved to be valid.

\subsection{how do we interpret the divergence in a wave function?}
It does not make sense to interpret the divergence directly by probability which has been proven by various experiments, but the essence remains unclear\cite{AdlerBassi-24}. Comparing the eigenfunction of the divergent state and the convergent state, the biggest distinction is that the former is divergent, and the latter is convergent, and meanwhile, one of them is stable, and the other is unstable. Thus, we can comfortably conclude that the divergent wave function at least suggests the state is unstable. It should experience two possible processes from a divergent unstable state to a convergent stable state. (\romannumeral1) The energy remains unchanged, and the divergent state will self-tune to reach the more stable state, which is achieved by adjusting the coefficient $B_0$(or $A_0$) to increase the size of $\Delta{x}$, which happens at the $\alpha$ state; (\romannumeral2) The state exchanges energy with the outside to reach a more stable state, which is achieved by adjusting the energy to get a bigger $\Delta{x}$, which happens at the $\beta$ state. A divergent unstable state will experience self-tuning and exchange energy with the outside to reach a convergent stable state. It is obvious that the size of $\Delta{x}$ is closely related to the stability of the state.

\begin{figure}
\begin{center}
\includegraphics [width=4.5in]{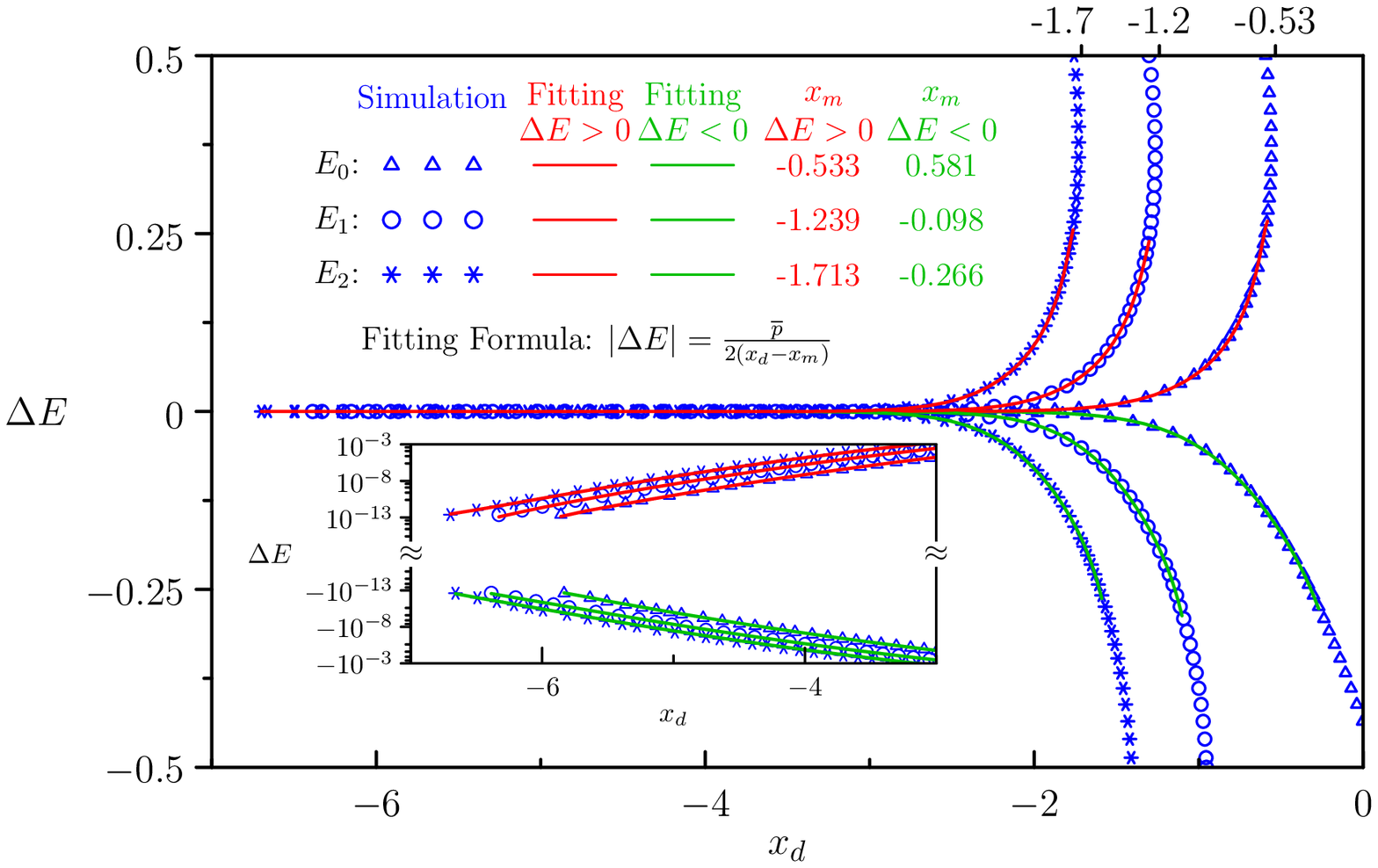}
\caption{\textbf{Schematic view of energy distribution.} The green-filled region schematically indicates the form of energy distribution, which does not directly mean the amount of energy for it has not been squared. To illustrate this clearly, the amount of energy to exchange is amplified. With a natural unit, the width of the potential well is 6, and the height of the potential is 8.}\label{fig3}
\end{center}
\end{figure}

We know that the total probability of the divergent part is $\frac {\Delta{E}}{E}$ or $\frac {\Delta{E}}{E_n}$, which is obviously limited. The $\beta$ state should exchange a part of its energy with the outside to reach a stable state. meanwhile, the energy distributed in the divergent part is $\Delta{E}$. For a normalized wave function, such as in figure \ref{f1}, the divergence is still there, and the integration of $\oint \Psi^* \hat{H}\Psi \,\rm d x$ in the whole divergent part is infinite. This is unacceptable. The normalized wave function should not be renormalized in spite of the divergence. Moreover, integration of $\oint \Psi^* \hat{H}\Psi \,\rm d x$ should be $\Delta{E}$. Thus, the integration of the divergent part of the wave function should be done over the corresponding interval, namely, $\oint_c \Psi^* \hat{H} \Psi \,\rm d x=\Delta{E}$. The energy density $\Psi^* \hat{H}\Psi$ changes with space. The larger $\Psi(x)$ is, the bigger the corresponding energy density is. In the area of minus infinity, $\Psi(-\infty)$ is infinitely great, such as in figure \ref{f1}, and its integration interval should be infinitely small, such as shown in figure \ref{fig3}. As we know, the normal part of a wave function is a standing wave where the energy is not transmitting. From the angle of the integration interval, the divergent part of a wave function is a traveling-like wave, which the energy is transmitting. We argue that the divergent part of the wave function denotes the amplitude path or track of the divergent traveling wave. As shown in figure \ref{fig3}, for the states of $\psi_{E<E_0}$, all of its energy is distributed around the potential well. For the states of $\psi_{E\ge E_0}$, the small energy block distributed away from the potential well is the photo taken at different times. This part is traveling outward, and the distribution interval is adjusted in real time to keep the total energy of the divergent part unchanged. Interpretation of the divergent with the integral interval can be found in ref\cite{HatanoKawamoto-69} where the interval is a whole block.

Energy density $\phi^*\hat{H}\phi$ is just the probability $\phi^*\phi$ times energy $E$. This is identical to the description of rules for quantum electrodynamics. If it is true that the concept of energy distribution can be generalized to a particle, the physical meaning behind it will be interesting. Energy is detectable. The bigger the energy density is, the greater the signal received by the detector, which is treated as probability. Energy distribution not only eliminates the divergence difficulty, but also provide a connection between the wave function and a measurable physical quantity, namely, probability. The process of normalization is to guarantee the conservation of energy.

\subsection{what is the difference between two different diverging directions?}
Up to now, the meaning of diverging direction was clear. Namely, it suggested the state was going to absorb or release energy. Along with the energy change, when the $\pm\beta$ state transforms into the $\mp\beta$ state, then there certainly is a convergent stable state between them. With self-tuning and exchange energy with the outside, the wave functions at the boundary will spontaneously adjust to zero or other patterns of convergence which are the most stable. The self-regulating mechanism illuminates why the boundary condition is what it is. The essence of self-tuning and energy exchange is the adjustment of an energy distribution which is related to the stability of the state. 

The divergent state is unstable. To achieve state stability, the energy distribution will be adjusted by self-tuning and energy exchange. This self-adjustment is realizable in simulation. Behind the behavior, the wave function should own a mechanism to adjust the distribution of energy. Clearly, the divergent wave function carries sufficient information to reach the stable state efficiently, and the process is spontaneous. Speaking plainly, a divergent state knows how to reach a stable state. Namely, it knows: (\romannumeral1) whether it ought to absorb or release energy; (\romannumeral2) the exact amount of energy to exchange, and (\romannumeral3) how to adjust the energy distribution to optimal.

\subsection{Divergence criterion}
It has been already proved that $\Delta{E}$ is related to $\Delta{x}$. The energy minimization principle is often used to obtain the ground state. It corresponds to $\Delta{E}=0$, while $\Delta{x}$ corresponds to the diverging extent of the wave function which is related to the stability. These considerations make another criterion possible. Namely, whether the wave function is divergent is the criterion for whether the particle has reached a steady state, and the precondition is that all continuity conditions are met well. The extent of divergence determines the uncertainty of the energy, namely the stability of the state. Then, in practice, the extent of divergence determines the energy accuracy of a selected stable state. It is obvious that this criterion is also suitable for excited states. Moreover, it is absolutely reliable, because the stability of the wave functions is directly determined by $\Delta{x}$. The bigger that $\Delta{x}$ is, the more stable the state is. As long as there is no divergence, $\Delta{x}$ is naturally big enough. Then the stability is naturally guaranteed. Under the circumstances, this criterion has a wide range of applications, such as in thermodynamic simulation\cite{IbanezNezgaraot-195}, the many-body problem, time-dependent problems, etc. Once one achieves a convergent wave function in any way that satisfies the precondition, one has obtained a convergent stable state.

\subsection{1D Quantum Shooting Method}
Now we can refine the shooting method we used in the beginning. In a normal shooting method, the diverging direction is used to adjust the direction of shooting\cite{GiordanoNakanishi-32,PressTeukolsky-33}. It is used as a numerical method without exploring the physics underneath. As in the above statement, the diverging direction and extent are described by the non-hermitian energy-space uncertainty relation. $\Delta{x}$ can be read directly from the wave unction, and then the approximate value of $\Delta{E}$ is obtainable through the uncertainty relation. Then we have a more accurate energy of the nearby convergent stable state. The energy-space uncertainty relation serves as the gun sight. It is much more efficient than the diverging direction used alone. We contend that this refined shooting method is not only a numerical method, but also comprises the actual physical images.

\section{Conclusions}
Unlike conventional methods of utilizing the hermiticity or the $PT$-symmetry of a Hamiltonian to guarantee the reality of its eigenvalue, we set the eigenvalue of the Hamiltonian to be real at first. Meanwhile, the open and semi-open boundary conditions were adopted. The consequent divergence can be described well by the energy-space uncertainty relation. The mechanism of energy exchange in a divergent unstable state is just in the uncertainty relation, which supports the energy density as being associated with the probability density, which benefits the controversy involving the essence of probability. The divergent wave function is normalizable with energy density. A divergent criterion, which is analogous to the energy minimization principle in conventional quantum mechanics, is presented to determine whether the state is stable or not. In short, divergent states and the physical laws under their regularities provide us a new strategy to overcome difficulties in the quantum field, especially the many-body problem. Based on our research, the wave function of a particle is the same as a wave function of an electromagnetic field, both of which share the conclusion that the square of the wave function denotes the energy density. It is apparent in the situation of an electromagnetic field, but as for a particle, it is hard to accept, which deserves further deeper research.

\section*{Acknowledgment}
The autor gratefully acknowledges the financial support of the Xiangtan University start-up Foundation (No. KZ08088).
\section*{References}


\end{document}